\documentclass[conference]{IEEEtran}

\ifCLASSINFOpdf
  
\else
  
\fi
\usepackage[letterpaper,bindingoffset=0.1in,%
            left=1in,right=1in,top=0.75in,bottom=1in,%
            footskip=.25in]{geometry}
            	
\usepackage{amsfonts}

\usepackage{adjustbox}
\usepackage{algorithmicx}
\usepackage{algpseudocode}
\usepackage[ruled]{algorithm}
\usepackage{graphicx}
\usepackage{caption}
\usepackage{graphicx,times,amsmath} 
\usepackage{caption}
\usepackage{subcaption}
\usepackage[rightcaption]{sidecap}
\usepackage{caption}
\usepackage{multirow}
\usepackage{hhline}
\usepackage{amsmath}
\usepackage{epstopdf}
\usepackage{tabularx}
\usepackage{fancyhdr}

\graphicspath{ {images/} }
\IEEEoverridecommandlockouts \IEEEpubid{\makebox[\columnwidth]{ 978-1-5386-3531-5/17/\$31.00~\copyright~2017 IEEE \hfill} \hspace{\columnsep}\makebox[\columnwidth]{ }}

\usepackage{fancyhdr}
\fancypagestyle{pageStyleOne}{%
\normalfont\footnotesize
    \fancyhf{}
\fancyhead[L]{\fontsize{8}{8} \selectfont This paper is accepted for presentation at 2017 IEEE 28th Annual International Symposium on Personal, Indoor, and Mobile Radio Communications.}
}

\makeatletter

\begin{document}


\title{A Convolutional Neural Network for\\ Search Term Detection}

\author{
    \IEEEauthorblockN{Hojjat Salehinejad\IEEEauthorrefmark{1}\IEEEauthorrefmark{2}, 
    Joseph Barfett\IEEEauthorrefmark{2}, 
    Parham Aarabi\IEEEauthorrefmark{1}, 
    Shahrokh Valaee\IEEEauthorrefmark{1}, \\
    Errol Colak\IEEEauthorrefmark{2},
    Bruce Gray\IEEEauthorrefmark{2}, and
    Tim Dowdell\IEEEauthorrefmark{2}}
    \IEEEauthorblockA{\IEEEauthorrefmark{1}Department of Electrical \& Computer Engineering, University of Toronto, Toronto, Canada}
        \IEEEauthorblockA{\IEEEauthorrefmark{2}Department of Medical Imaging, St. Michael's Hospital, University of Toronto, Toronto, Canada }

           \{salehinejadh, barfettj\}@smh.ca, p@arh.am, valaee@ece.utoronto.ca,  \{colake, grayb, dowdellt\}@smh.ca
       
}


\maketitle

\thispagestyle{pageStyleOne}

\begin{abstract}
Pathfinding in hospitals is challenging for patients, visitors, and even employees. Many people have experienced getting lost due to lack of clear guidance, large footprint of hospitals, and confusing array of hospital wings. In this paper, we propose Halo; An indoor navigation application based on voice-user interaction to help provide directions for users without assistance of a localization system. The main challenge is accurate detection of origin and destination search terms. A custom convolutional neural network (CNN) is proposed to detect origin and destination search terms from transcription of a submitted speech query. The CNN is trained based on a set of queries tailored specifically for hospital and clinic environments. Performance of the proposed model is studied and compared with Levenshtein distance-based word matching. 
\end{abstract}


\section{Introduction}
Medical environments such as hospitals host a large number of admitted patients, out-patients, visitors, and employees. Many of these individuals are one-time or short-term visitors that often have difficulty in finding their ultimate destination \cite{guardian}. Hospitals have implemented potential solutions in the form of paths on the floors or walls, installation of signs, and directory finder systems on their websites. Unfortunately, these approaches have not fully solved the problem and these aids are not available at every point of a hospital \cite{guardian}. Statistics show that in UK around 6.9M outpatient hospital appointments are missing annually, where a significant fraction of these is due to navigation problems inside big hospitals \cite{guardian}. 

Artificial intelligence have been used successfuly in different navigation applications such as vehicle navigation \cite{salehinejad2010dynamic} and 3-dimensional localization \cite{salehinejad20143d}. The main challenges in providing a reliable indoor navigation solution for hospital environments are design of a user-friendly interface and an accurate indoor localization system. The focus of this paper is the user-application interaction component of such a system, without an integrated indoor localization system. 

The voice-user interaction (VUI) systems are typically equipped with speech-to-text (STT) and text-to-speech (TTS) engines that employ machine learning models to convert speech to text and vice versa. To be useful in the context of navigation in a hospital, the application will be required to detect the important keywords and/or sentiment from the user query. For our indoor navigation application, these keywords are the current location and ultimate destination of user.

Deep learning has achieved significant performance in many domains, mainly in natural language processing (NLP) \cite{graves2013speech} and machine vision \cite{krizhevsky2012imagenet} problems. Convolutional neural networks (CNN) are one of the fundamental deep learning models with translation invariance characteristics that requires minimal preprocessing of input data. Some applications of NLP are sentence modeling \cite{blunsom2014convolutional} and sentence classification \cite{kim2014convolutional}. 

In this paper, we present Halo; A user-friendly navigation application targeted for hospital environments without localization systems assistance. We propose a task specific CNN for detection of start point and destination in submitted user queries. Instead of a single softmax output layer \cite{kim2014convolutional}, our model uses two distinct softmax layers for origin and destination search term prediction. The custom CNN model is trained using department names and key locations within a hospital. Our key contributions include:

- A generated labeled dataset of direction seekers queries based on medical department keywords. 

- Design of a customized CNN, trained with medical department keywords for navigation inside hospitals. 

- Application design and implementation of an interface for indoor navigation without localization assistance.

A brief overview on recent keyword detection and learning vector space representation is provided in Section~\ref{sec:back}. The Levenshtein distance is discussed in Section~\ref{sec:lev}. Section~\ref{sec:model} presents the proposed model and the experimental results are discussed in Section~\ref{sec:exp}. The paper is concluded in Section~\ref{sec:conc}.

\section{Background}
\label{sec:back}
CNN are trained to fill out a frame of slot-value pairs for sub-dialog segments based on a pool of dialog topics. A multi-topic detection model can have a convolutional layer with three feature maps: one for each input topic and one shared between the two inputs \cite{shi2017convolutional}. The CNN model can work on character to sentence level information exploration to perform sentiment analysis of short texts such as single sentences and Twitter messages \cite{dos2014deep}. For larger inputs with a large vocabulary dictionary, machine learning models generally use a dimensionally reduced representation of text.

\begin{figure}[t]
\centering
\captionsetup{font=footnotesize}
                \includegraphics[width=0.22\textwidth]{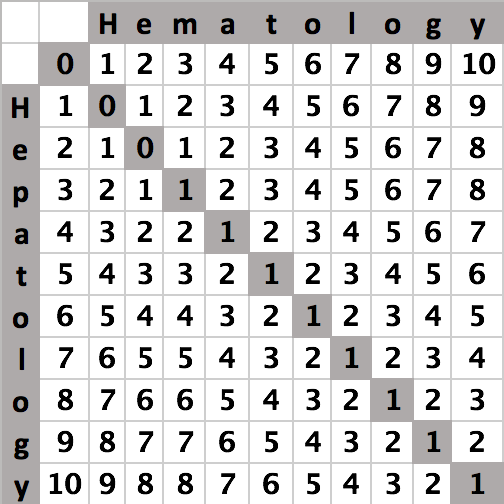}
\caption{Levenshtein distance for ``Hepatology" and ``Hematology".}
\label{fig:leven}
  \vspace{-1.5em}
\end{figure}

Word2vec is a learning vector space representation of words, which produces a vector space from a large corpus of text with much less dimensionality. This model assigns a word vector to each unique word in the corpus. The vectors with closer context are positioned in closer proximity in this space \cite{mikolov2013distributed}. 
The word2vec can use continuous bag-of-words (CBOW) or continuous skip-gram for dimension reduction \cite{mikolov2013distributed}, \cite{mikolov2013efficient}. The CBOW provides a bag of words in which the order of the words is not important and can predict the current word from a window of surrounding context words from the bag. The continuous skip-gram considers heavier weights for the surrounding context words of the current word; however, it is slower for infrequent words \cite{mikolov2013efficient}. GloVe provides a global vector space representations of words. It uses global matrix factorization and local context window methods to construct a global log-bilinear regression model \cite{pennington2014glove}. The matrix of word-word co-occurrence is sparse by nature. However, GloVe only uses the non-zero elements for training and even do not consider individual context windows in a large corpus \cite{pennington2014glove}. The CNN work well on $n$-gram representation of input data \cite{majumder2017deep}. A convlution layer can perform feature extraction from various $n$ values of a $n$-gram model and perfrom personality detection from documents \cite{majumder2017deep}. Dynamic $k$-Max Pooling operates as a global pooling operation over linear sequences, suitable for semantic modelling of sentences \cite{kalchbrenner2014convolutional}. This model receives variable dimension size input sentences and induces a feature graph over the sentence that is capable of explicitly capturing short and long-range relations \cite{kalchbrenner2014convolutional}.

A CNN with a single output softmax layer can classify a vector representation of text with high accuracy \cite{kim2014convolutional}. The design of output layer in CNN can vary depending on the application of the model. Recurrent neural networks (RNN) can learn long-term dependencies \cite{salehinejad2016learning} over sequential data such as in genomics \cite{pouladi2015recurrent} and shopping pattern of customers throught time \cite{salehinejad2016customer}. A collaboration of RNN and CNN for feature extraction of sentences is possible. A CNN learns features of an input sentence and then a gated RNN model discourse information \cite{ren2017neural}. A bi-directional long short term memory (BLSTM) RNN can sequentially read words from question and answer sentences for answer sentence selection and to produce a corresponding score \cite{Wang2015ALS}.

\begin{figure}[!tbp]
\centering
\captionsetup{font=footnotesize}
\includegraphics[width=0.4\textwidth]{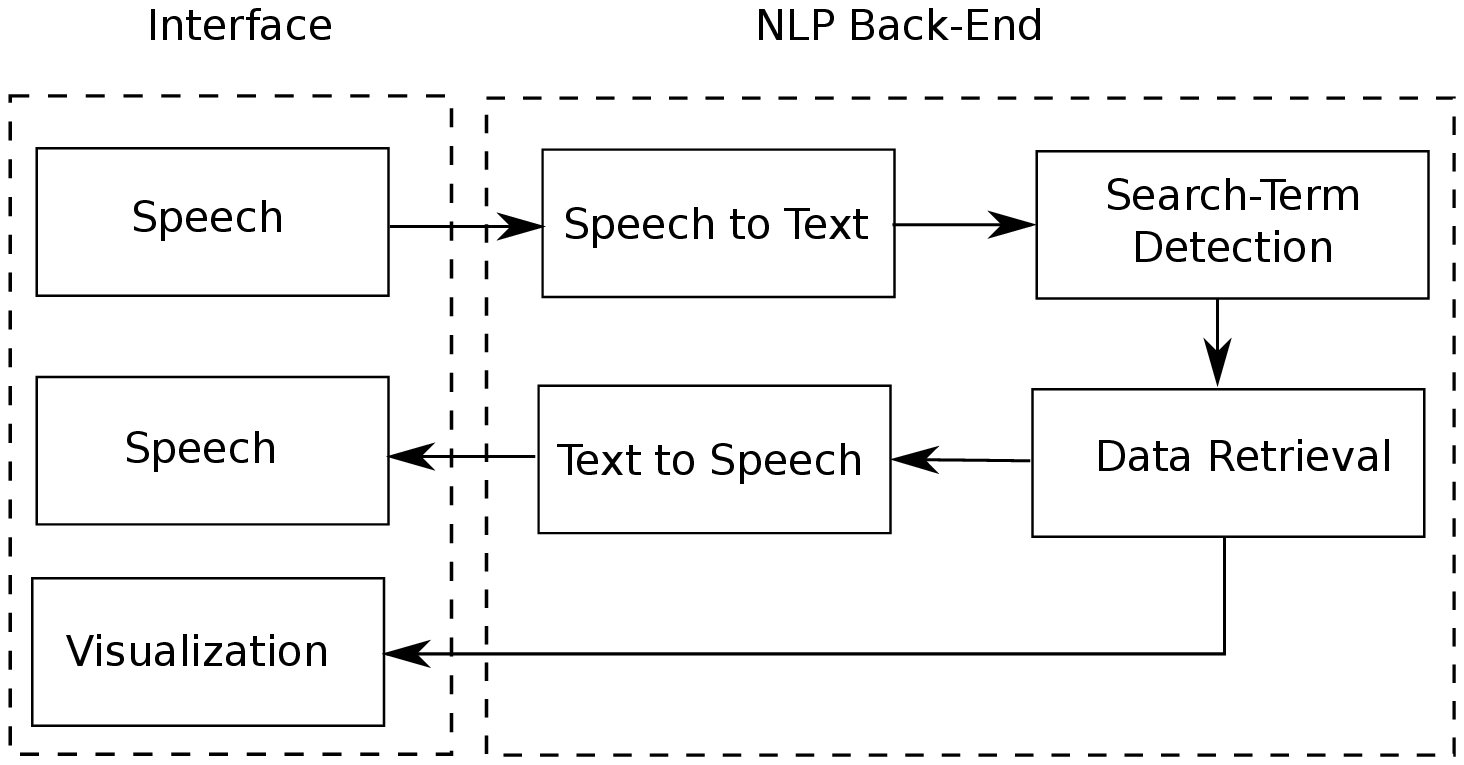}
\caption{Proposed model architecture.}
\label{fig:model}
  \vspace{-1.5em}
\end{figure}

\begin{figure*}[!htbp]
\centering
\captionsetup{font=footnotesize}
                \includegraphics[width=0.7\textwidth]{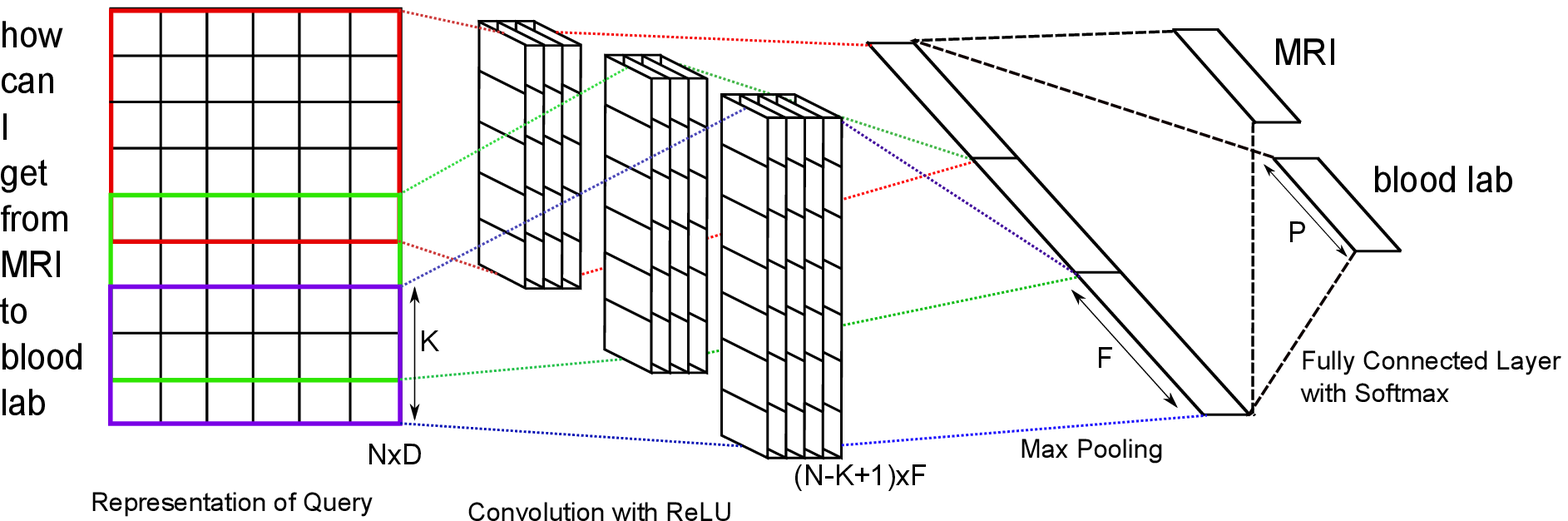}
\caption{Proposed convolutional neural network for search-terms detection from submitted query.}
\label{fig:cnn}
  \vspace{-1.5em}
\end{figure*}

\section{Levenshtein Distance}
\label{sec:lev}
The Levenshtein distance, also known as the ``edit" distance, refers to the smallest number of insertions, deletions, and substitutions required to change one string into another \cite{navarro2001guided}. It is used in TextRank for automatic keyword extraction 
and summarization \cite{mihalcea2004textrank}. For two strings $s_{1}$ and $s_{2}$ with lengths $|s_{1}|$ and $|s_{2}|$, respectively, $\mathcal{L}(s_{1}^{i},s_{2}^{j})$ is defined as the distance between the first $i$ characters of $s_{1}$ and first $j$ characters of $s_{2}$ such that

\begin{equation}
\mathcal{L}(s_{1}^{i},s_{2}^{j})=
 \begin{cases}
\mathcal{L}(s_{1}^{i-1},s_{2}^{j-1})  & s_{1}^{i}=s_{2}^{j}\\    
1+ L & otherwise\    
\end{cases}
\end{equation}
where 
\begin{equation}
L = min(\mathcal{L}(s_{1}^{i-1},s_{2}^{j}), \mathcal{L}(s_{1}^{i},s_{2}^{j-1}), \mathcal{L}(s_{1}^{i-1},s_{2}^{j-1}))
\end{equation}
and $\mathcal{L}(s_{1}^{i},s_{2}^{0})=i$ and $\mathcal{L}(s_{1}^{0},s_{2}^{j})=j$; $s_{k}^{0}$ represents an empty string. In a matrix representation, $\mathcal{L}(s_{1}^{i},s_{2}^{j})$ is the minimum number of operations needed to match $s_{1}^{i}$ and $s_{2}^{j}$. The Levenshtein distance for the strings ``Hepatology" and ``Hematology" is presented in Figure~\ref{fig:leven}.

For the navigation case study in hospitals and medical centres, one needs to compute the Levenshtein distance between each recognized word in $V$ with a list of keywords $D$ corresponding to the directories in the hospital, that is $|V|\times |D|$ times.

The next step is deciding on the origin and the destination in the detected sentence. The typical query to ask for direction submitted by users follows a patterns such as

\textit{``how can I get \textbf{from} $\underbrace{cardiology}_\text{origin}$ \textbf{to} $\underbrace{hematology}_\text{destination}$?"} \\
or

\textit{``how can I get \textbf{to} $\underbrace{hematology}_\text{destination}$ \textbf{from} $\underbrace{cardiology}_\text{origin}$?".}

The first example shows that the origin and destination search terms are occurring in the order of origin-destination (i.e. origin is \textit{cardiology} and destination is \textit{hematology}). However, the origin and destination can be inverted as in the second phrase. Deterministic methods, including Levenshtein distance, require hand-crafted patterns to detect origin and destination in a sentence. However, such effort is not efficient enough, as queries might be submitted in other patterns or words with similar pronunciation can be detected in different ways such as ``two" or ``too" instead of  ``to". Therefore, we need a solution to consider the dependency among words in a sentence and understand different patterns of queries.

\section{Model Architecture}
\label{sec:model}
Architecture of the proposed model is presented in Figure~\ref{fig:model}. This architecture consists of two environments, which are the interface and natural language processing (NLP) back-end. The interface is responsible for receiving a speech command from user and returning the output from the NLP back-end to the user, both visually and audibly. The NLP back-end consists of four main components, which are STT, search-term detection, data retrieval, and TTS engines. Our focus in this paper is on the search-term detection detection component. 

\subsection{Speech to Text} 
The STT refers to the process of converting the received speech signal to text. Following the rapid advances and proliferation of NLP, many STT models and application program interfaces (API) are now available for use. The back-end in Figure~\ref{fig:model} sends the received speech signal to the Google STT API. The search-term detection engine then receives the transcribed speech as a text vector $V$.

\subsection{Convolutional Neural Networks}
\label{sec:cnn}

In this section, we propose a sentence-level-classification task specific model for search-term detection (i.e. origin-destination) based on CNN, presented in Figure~\ref{fig:cnn}. 

A preprocessed query coming from the STT engine with $N$ words is represented as a sequence
$\textbf{V}= [\textbf{v}_{1},... ,\textbf{v}_{N}]$. Each word can be represented in a word-vector form as $\textbf{x}_{n}\in \mathbb{R}^{D}$ where $D$ is the vector dimension. This word-vector representation is an outcome of projecting each word from a sparse vector space of dimension $M$ to a lower dimension $D$ such that $M>D$. In the latter space, the semantically close words are closer to each other. In this paper, we have used word2vec, which was pre-trained using 100 billion words from Google News \cite{kim2014convolutional}, \cite{mikolov2013distributed}.

A concatenated representation of the vector $V$ for words $1$ to $N$ is presented as
\begin{equation}
\textbf{x}_{1:N}=\textbf{x}_{1}\oplus \textbf{x}_{2}\oplus...\oplus \textbf{x}_{N}
\end{equation}
where $\oplus$ is the concatenation sign.

As Figure~\ref{fig:cnn} shows, the model has three filters with varying window sizes $K\in\{3,4,5\}$ that slide across the 
input layer \cite{kim2014convolutional}. The filters extract features from the input layer to construct feature maps of size $(N-K+1)\times F$, where $F$ is the number of feature maps for each filter. Each feature map $\textbf{h}_{1 \times (N-K+1)}$ has its own shared weight $\textbf{w}_{K\times D}$ and bias $\textbf{b}_{1\times N-K+1}$. The value of a hidden neuron $m$ is

\begin{equation}
h_{m}=\sigma(\sum_{k=1}^{K}\sum_{d=1}^{D} x_{m+k-1,d} \cdot w_{k,d} +b_{m})
\end{equation}
where the window word is $\textbf{x}_{m:m+K-1}$ and $\sigma(\cdot)$ is a rectified linear unit (ReLU) activation function defined as 
\begin{equation}
\sigma(z)=max(0,z)
\end{equation} 
where $z\in\mathbb{R}$.

The max pooling over time \cite{collobert2011natural} extracts the most significant feature from the feature map of a particular filter such as
\begin{equation}
\hat{h}_{l}=max(\{h_{1},..,h_{N-K+1}\}).
\end{equation}

The output of the max-pooling layer contains the max-pooled features $\hat{\textbf{h}}=[\hat{h}_{1},...,\hat{h}_{L}]$ where the length of features vector is $L=3F$. These features are passed to a fully connected perceptron network. During training, the drop-out regularization randomly drops units along with their connections from the 
input layer using a binary \textit{mask} vector $\textbf{r}_{1\times L}$ with Bernoulli random distribution (with probability $p=1$). 

The output layer has two separate softmax output layers; One for the origin search-term detection and one for the destination search-term detection. The output of each layer is probability distribution over all labels (departments) for origin and destination. Therefore, in case of having $P$ distinguish departments, the output layer totally has $2\times P$ units. The value of label $p$ is predicted as

\begin{equation}
y_{p} = \phi(\sum_{l=1}^{L} (\hat{h}_{1,l} \cdot r_{1,l})\cdot w_{l,p} + b_{p})
\end{equation}
where $w_{l,p}$ is the weight of connection from feature $l$ to label $p$ and $\phi(\textbf{z})$ is the softmax activation function defined as
\begin{equation}
\phi(z_{p})=\frac{e^{z_{p}}}{\sum_{j=1}^{P}e^{z_{j}}} \:\: for \:\: p=[1,...,P].
\end{equation}

The advantage of having a dedicated output layer for each search term is fewer number of output units (i.e. $2P$) comparing to a single output layer model where each output unit represent a search term pair and has $P(P-1)$ labels. The number of optimization parameters in the output layer of first and second models is $2P(3F+1)$ and $P(P-1)(3F+1)$, respectively. As the number of search terms $P$ increases, the complexity of output layer grows linearly and exponentially for the dedicated and single output layer models, respectively. 

\subsection{Text to Speech} 
The input to the TTS engine is the words vector $T$, representing the navigation instructions to the user.
This vector is passed to a TTS engine to convert it to speech. 

\section{Experiments}
\label{sec:exp}
Halo is implemented in Kivy and Python. Kivy is a GPU accelerated framework available for multiple operating systems. 
The speech to text component uses Google STT (Speech-to-Text) API \cite{google} and the text to speech component uses the playsound library in Python. The CNN model is implemented in Tensorflow \cite{abadi2016tensorflow}.

We compare the CNN model with random forest (RF) \cite{breiman2001random} and support vector machine (SVM) \cite{smola2004tutorial}. Instead of using a bag-of-the-words technique, which considers frequency of the words in a query as the features, a $n$-gram model with a lower boundary $n=1$ and upper boundary $n\in\{1,2,3\}$ performs the feature extraction of raw queries for SVM and RF models. The $n$-gram model stores the spatial information within the text depending on the value of $n$. For example, for the query\\
 \textit{``I want to go to \textbf{Trauma Clinic} from \textbf{Rapid Referral Clinic}."}\\
 the bi-gram (i.e. 2-gram) representation is 
[I, want, to , go , to, Trauma, Clinic, from, Rapid, Referral, Clinic, I want, want to, to go, go to, to Trauma, Trauma Clinic, Clinic from, from Rapid, Rapid Referral, Referral Clinic]. This example shows that bi-gram model detects the search term ``Trauma Clinic" as a gram but  the search term ``Rapid Referral Clinic" is not detected as a unique gram  \cite{majumder2017deep}.  A hashing function maps the grams to indices.

\subsection{Data}
We have built a dataset to train Halo and evaluate its performance. For 79 medical departments, 283,452 unique queries are generated based on most common English queries used for asking for a direction. Some of the examples are

\textit{``I want to go to \textbf{Trauma Clinic} from \textbf{Rapid Referral Clinic}."}\\
and 

\textit{``I have to go to \textbf{MRI} from \textbf{Mental Health Services}."}

The Halo dataset will be available online for research purposes\footnote{www.mimlab.ca}.


\subsection{Settings}
The proposed CNN is simple, has small number of hyper-parameters and is reasonably easy to train.          
It is trained with a mini-batch size of 64, drop-out probability of 0.5, filter sizes 3, 4, 5 and 100 feature maps per filter size. The number of training iterations is set to 10 and the Adam optimizer with the learning rate of 0.001 is used \cite{kingma2014adam}. The weights at output layers are initialized using Xavier initializer \cite{glorot2010understanding}, the weights in convolutional layer are selected based on normal distribution with the standard deviation of 0.1, and biases are set to 0.1. The activation function before the max-pooling layer is ReLU \cite{nair2010rectified}. The $L_{2}$ regularization is set to $1.0 \times 10^{-4}$ and early-stopping is applied.

We use an implementation of RF \cite{breiman2001random} which combines classifiers by averaging their probabilistic prediction. Number of estimators is set to 10. The SVM model has a penalty parameter of one with radial basis function kernel.  

For all the experiments, $70\%$ and $30\%$ of data is allocated for training and testing, respectively.  The data is shuffled before splitting into training and testing datasets and the results after 10-fold cross validation are reported.

\subsection{Performance Evaluation}
Halo is able to communicate with a user and visually demonstrate the current location and path to follow to the destination.
A user can press the ``Tap to Speak" button to talk to Halo. When the direction is found by Halo, it will respond by displaying a floor map with directions and speech commands. 

 As an example, a user can submit the following query in the form of speech to the application:

\textit{``How can I get from \textbf{reception} to \textbf{MRI}?"}\\
where the application responds back as

\textit{``From reception turn right. Walk along the hallway and turn right at the first turn. You will see the sign for MRI. Walk along the hallway and turn left at the second turn. The MRI is at the your right."}.

\begin{table}[]
\centering
\caption{Matching accuracy in detection of origin and destination pairs on test dataset by Levenshtein distance (LD), random forests (RF), support vector machine (SVM), and convolutional neural networks (CNN) on Halo dataset, after 10-fold cross-validation.}
\label{T:result}
\begin{adjustbox}{width=0.49\textwidth}

\begin{tabular}{|c|c|c|c|}
\hline
Model                & Origin Detection & Destination Detection & Total   \\ \hline
LD& \%58.50\%          & 27.04\%               & 42.77\% \\ \hline
RF - 1-gram                 & 21.83\%          & 21.80 \%              & 21.81\% \\ \hline
RF - 2-gram                 & 92.66\%         & 92.60 \%              & 92.63\% \\ \hline
RF - 3-gram                 & 95.83\%          & 95.45\%               & 95.64\% \\ \hline
SVM - 1-gram                 & 22.55\%          & 23.41\%               & 22.98\% \\ \hline
SVM - 2-gram                 & 89.56\%          & 89.20\%               & 89.38\% \\ \hline
SVM - 3-gram                 & 92.21\%          & 92.30\%               & 92.26\% \\ \hline
CNN                  & \textbf{99.99\%}         & \textbf{99.99\%}              &\textbf{ 99.99\%} \\ \hline
\end{tabular}
\end{adjustbox}

\end{table}

The matching performance results of Levenshtein distance, RF \cite{breiman2001random},  SVM \cite{smola2004tutorial}, and CNN are presented in Table~\ref{T:result}. The overall performance of Levenshtein distance model is $42.77\%$. The RF and SVM models have competitive performance. The RF with 3-gram has better performance that RF 1-gram and RF 2-gram models with a matching accuracy of $95.64\%$. The SVM model with 3-gram feature extraction has a slightly less detection accuracy than the RF model with 3-gram. The proposed CNN model reaches $99.99\%$ matching accuracy after 8 training iterations to match the input inquiry to the correct origin-destination pair for both origin and destination search terms. 

As an example, for the query

\textit{``I want to go to \textbf{Admitting} from \textbf{Fracture Clinic}."}\\
the Levenshtein distance method compares each word of the query with the words in its dictionary. The prediction for each word of query is presented in Table~\ref{T:prediction}.
Levenshtein distance can detect the ``Admitting" and ``Fracture Clinic" correctly; However, it has also detected three other terms with the shortest Levenshtein distance to the word ``clinic", which are ``MRI Clinic", ``Eye Clinic", and ``Spine Clinic".  Three terms are detected instead of two as the origin-destination terms. On the other hand, the CNN approach can learn the dependency among the words in a query and represent the origin and destination accordingly; Such that origin is ``Fracture Clinic" and destination is ``Admitting" in this example.

\begin{table}[]
\centering
\caption{Prediction(s) of search-terms in input query \textit{``I want to go to \textbf{Admitting} from \textbf{Fracture Clinic}"}  by Levenshtein distance (LD) and CNN.}
\begin{adjustbox}{width=0.49\textwidth}
\begin{tabular}{|c|c|c|}
\hline
Original  & Prediction(s) by LD  & Prediction by CNN \\ \hline
I         & -                                      & -                 \\ \hline
want      & -                                      & -                 \\ \hline
to        & -                                      & -                 \\ \hline
go        & -                                      & -                 \\ \hline
to        & -                                      & -                 \\ \hline
Admitting & Admitting                              & Admitting         \\ \hline
from      & -                                      & -                 \\ \hline
Fracture  & Fracture Clinic                        & Fracture   \\ \hline
Clinic    & MRI Clinic - Eye Clinic - Spine Clinic &          Clinic         \\ \hline
\end{tabular}
\end{adjustbox}
\label{T:prediction}
\end{table}

\section{Conclusion and Future Works}
\label{sec:conc}
This paper presents system design and implementation for indoor navigation inside medical buildings such as a hospital. A training set for machine learning applications is built. A customized convolutional neural network is designed and trained based on provided open-source data for origin-destination search terms detection from submitted queries. The results show that the proposed CNN has high performance in detecting the origin and destination search terms, regardless of the input query pattern or the voice of user. 

The CNN search term detection model can be utilized for voice control applications in blind navigation systems. It also can directly work on the speech query without need for conversion to text. This approach is not limited to hospitals and can be deployed for navigation without localization assistance in shopping centers and on university campuses, to mention some. It is interesting to implement a recurrent neural network (RNN) as another machine learning approach.
These models need to be evaluated for ambiguous ascent from complicated backgrounds and when grammatically incorrect queries are given as input.

\bibliographystyle{IEEEtran}
\bibliography{CTLIEEEtrans,mybibfile}

\begin{thebibliography}{10}
\providecommand{\url}[1]{#1}
\csname url@samestyle\endcsname
\providecommand{\newblock}{\relax}
\providecommand{\bibinfo}[2]{#2}
\providecommand{\BIBentrySTDinterwordspacing}{\spaceskip=0pt\relax}
\providecommand{\BIBentryALTinterwordstretchfactor}{4}
\providecommand{\BIBentryALTinterwordspacing}{\spaceskip=\fontdimen2\font plus
\BIBentryALTinterwordstretchfactor\fontdimen3\font minus
  \fontdimen4\font\relax}
\providecommand{\BIBforeignlanguage}[2]{{%
\expandafter\ifx\csname l@#1\endcsname\relax
\typeout{** WARNING: IEEEtran.bst: No hyphenation pattern has been}%
\typeout{** loaded for the language `#1'. Using the pattern for}%
\typeout{** the default language instead.}%
\else
\language=\csname l@#1\endcsname
\fi
#2}}
\providecommand{\BIBdecl}{\relax}
\BIBdecl

\bibitem{guardian}
\BIBentryALTinterwordspacing
J.~Pinchin, ``Getting lost in hospitals costs the nhs and patients,'' 2015.
  [Online]. Available:
  \url{https://www.theguardian.com/healthcare-network/2015/mar/05/lost-hospitals-costs-nhs-patients-navigation}
\BIBentrySTDinterwordspacing

\bibitem{salehinejad2010dynamic}
H.~Salehinejad and S.~Talebi, ``Dynamic fuzzy logic-ant colony system-based
  route selection system,'' \emph{Applied Computational Intelligence and Soft
  Computing}, 2010.

\bibitem{salehinejad20143d}
H.~Salehinejad, R.~Zadeh, R.~Liscano, and S.~Rahnamayan, ``3d localization in
  large-scale wireless sensor networks: A micro-differential evolution
  approach,'' in \emph{Personal, Indoor, and Mobile Radio Communication
  (PIMRC), 2014 IEEE 25th Annual International Symposium on}.\hskip 1em plus
  0.5em minus 0.4em\relax IEEE, 2014, pp. 1824--1828.

\bibitem{graves2013speech}
A.~Graves, A.-r. Mohamed, and G.~Hinton, ``Speech recognition with deep
  recurrent neural networks,'' in \emph{Acoustics, speech and signal processing
  (icassp), 2013 ieee international conference on}.\hskip 1em plus 0.5em minus
  0.4em\relax IEEE, 2013, pp. 6645--6649.

\bibitem{krizhevsky2012imagenet}
A.~Krizhevsky, I.~Sutskever, and G.~E. Hinton, ``Imagenet classification with
  deep convolutional neural networks,'' in \emph{Advances in neural information
  processing systems}, 2012, pp. 1097--1105.

\bibitem{blunsom2014convolutional}
P.~Blunsom, E.~Grefenstette, and N.~Kalchbrenner, ``A convolutional neural
  network for modelling sentences,'' in \emph{Proceedings of the 52nd Annual
  Meeting of the Association for Computational Linguistics}.\hskip 1em plus
  0.5em minus 0.4em\relax Proceedings of the 52nd Annual Meeting of the
  Association for Computational Linguistics, 2014.

\bibitem{kim2014convolutional}
Y.~Kim, ``Convolutional neural networks for sentence classification,''
  \emph{arXiv preprint arXiv:1408.5882}, 2014.

\bibitem{shi2017convolutional}
H.~Shi, T.~Ushio, M.~Endo, K.~Yamagami, and N.~Horii, ``Convolutional neural
  networks for multi-topic dialog state tracking,'' in \emph{Dialogues with
  Social Robots}.\hskip 1em plus 0.5em minus 0.4em\relax Springer, 2017, pp.
  451--463.

\bibitem{dos2014deep}
C.~N. Dos~Santos and M.~Gatti, ``Deep convolutional neural networks for
  sentiment analysis of short texts,'' in \emph{COLING}, 2014, pp. 69--78.

\bibitem{mikolov2013distributed}
T.~Mikolov, I.~Sutskever, K.~Chen, G.~S. Corrado, and J.~Dean, ``Distributed
  representations of words and phrases and their compositionality,'' in
  \emph{Advances in neural information processing systems}, 2013, pp.
  3111--3119.

\bibitem{mikolov2013efficient}
T.~Mikolov, K.~Chen, G.~Corrado, and J.~Dean, ``Efficient estimation of word
  representations in vector space,'' \emph{arXiv preprint arXiv:1301.3781},
  2013.

\bibitem{pennington2014glove}
J.~Pennington, R.~Socher, and C.~D. Manning, ``Glove: Global vectors for word
  representation.'' in \emph{EMNLP}, vol.~14, 2014, pp. 1532--1543.

\bibitem{majumder2017deep}
N.~Majumder, S.~Poria, A.~Gelbukh, and E.~Cambria, ``Deep learning-based
  document modeling for personality detection from text,'' \emph{IEEE
  Intelligent Systems}, vol.~32, no.~2, pp. 74--79, 2017.

\bibitem{kalchbrenner2014convolutional}
N.~Kalchbrenner, E.~Grefenstette, and P.~Blunsom, ``A convolutional neural
  network for modelling sentences,'' \emph{arXiv preprint arXiv:1404.2188},
  2014.

\bibitem{salehinejad2016learning}
H.~Salehinejad, ``Learning over long time lags,'' \emph{arXiv preprint
  arXiv:1602.04335}, 2016.

\bibitem{pouladi2015recurrent}
F.~Pouladi, H.~Salehinejad, and A.~M. Gilani, ``Recurrent neural networks for
  sequential phenotype prediction in genomics,'' in \emph{Developments of
  E-Systems Engineering (DeSE), 2015 International Conference on}.\hskip 1em
  plus 0.5em minus 0.4em\relax IEEE, 2015, pp. 225--230.

\bibitem{salehinejad2016customer}
H.~Salehinejad and S.~Rahnamayan, ``Customer shopping pattern prediction: A
  recurrent neural network approach,'' in \emph{Computational Intelligence
  (SSCI), 2016 IEEE Symposium Series on}.\hskip 1em plus 0.5em minus
  0.4em\relax IEEE, 2016, pp. 1--6.

\bibitem{ren2017neural}
Y.~Ren and D.~Ji, ``Neural networks for deceptive opinion spam detection: An
  empirical study,'' \emph{Information Sciences}, vol. 385, pp. 213--224, 2017.

\bibitem{Wang2015ALS}
D.~Wang and E.~Nyberg, ``A long short-term memory model for answer sentence
  selection in question answering,'' in \emph{ACL}, 2015.

\bibitem{navarro2001guided}
G.~Navarro, ``A guided tour to approximate string matching,'' \emph{ACM
  computing surveys (CSUR)}, vol.~33, no.~1, pp. 31--88, 2001.

\bibitem{mihalcea2004textrank}
R.~Mihalcea and P.~Tarau, ``Textrank: Bringing order into texts.''\hskip 1em
  plus 0.5em minus 0.4em\relax Association for Computational Linguistics, 2004.

\bibitem{collobert2011natural}
R.~Collobert, J.~Weston, L.~Bottou, M.~Karlen, K.~Kavukcuoglu, and P.~Kuksa,
  ``Natural language processing (almost) from scratch,'' \emph{Journal of
  Machine Learning Research}, vol.~12, no. Aug, pp. 2493--2537, 2011.

\bibitem{google}
``https://cloud.google.com/speech/.''

\bibitem{abadi2016tensorflow}
M.~Abadi, A.~Agarwal, P.~Barham, E.~Brevdo, Z.~Chen, C.~Citro, G.~S. Corrado,
  A.~Davis, J.~Dean, M.~Devin \emph{et~al.}, ``Tensorflow: Large-scale machine
  learning on heterogeneous distributed systems,'' \emph{arXiv preprint
  arXiv:1603.04467}, 2016.

\bibitem{breiman2001random}
L.~Breiman, ``Random forests,'' \emph{Machine learning}, vol.~45, no.~1, pp.
  5--32, 2001.

\bibitem{smola2004tutorial}
A.~J. Smola and B.~Sch{\"o}lkopf, ``A tutorial on support vector regression,''
  \emph{Statistics and computing}, vol.~14, no.~3, pp. 199--222, 2004.

\bibitem{kingma2014adam}
D.~Kingma and J.~Ba, ``Adam: A method for stochastic optimization,''
  \emph{arXiv preprint arXiv:1412.6980}, 2014.

\bibitem{glorot2010understanding}
X.~Glorot and Y.~Bengio, ``Understanding the difficulty of training deep
  feedforward neural networks.'' in \emph{Aistats}, vol.~9, 2010, pp. 249--256.

\bibitem{nair2010rectified}
V.~Nair and G.~E. Hinton, ``Rectified linear units improve restricted boltzmann
  machines,'' in \emph{Proceedings of the 27th international conference on
  machine learning (ICML-10)}, 2010, pp. 807--814.

\end{thebibliography}

\end{document}